\documentclass[aps,prb,twocolumn, showpacs]{revtex4}

\usepackage{epsfig}
\usepackage{graphicx}

\def\be{\begin{equation}}
\def\ee{\end{equation}}
\def\bea{\begin{eqnarray}}
\def\eea{\end{eqnarray}}

\begin{document}
\newcount\timehh  \newcount\timemm
\timehh=\time \divide\timehh by 60
\timemm=\time
\count255=\timehh\multiply\count255 by -60 \advance\timemm by \count255

\title{Inelastic scattering approach to the theory of a tunnel magnetic transistor source}
\author{Emmanuel I. Rashba\cite{Rashba*} }
\affiliation{Department of Physics, SUNY at Buffalo, Buffalo, New York 14260, USA\\
and Department of Physics, MIT, Cambridge, Massachusetts 02139, USA}
\date{\today}
%\draft

\begin{abstract}
The high efficiency of a tunnel magnetic transistor as a source of spin-polarized electrons has been proven recently [X. Jiang {\it et al.}, Phys. Rev. Lett. {\bf 90}, 256603 (2003)]. A concept of this device based on an active group of hot electrons controlling the collector current and experiencing predominantly inelastic scattering in the base is developed. It takes into account the interconnection between the injection and filtering of spin-polarized electrons in different elements of the device. Explicit expressions for the parameters of the device in terms of the basic parameters of the system are presented. 
\end{abstract}

\maketitle

%\narrowtext

Despite spectacular progress achieved during the last couple of years, efficient spin injection from ferromagnetic (FM) metallic emitters still remains one of the central problems of semiconductor spintronics.\cite{reviews} There are two basic approaches to this problem. The first approach is based on the injection of nearly equilibrium electrons in a diffusive or a ballistic regime. It is well understood now that the failure of the initial attempts to inject spin polarized currents from FM metals into semiconductor microstructures was caused by ``perfect" (low resistance) contacts producing the ``bottlenecking"\cite{JS87} or ``conductivity mismatch"\cite{Sch00} problem. Since the role of resistive spin-selective contacts had been understood,\cite{R00} an impressive experimental progress has been achieved\cite{Plo01,HJ01,Jon01,Saf01,Weiss02,Jap02,Samarth02,Dutch02} and the attention of experimenters\cite{Han03} and theorists\cite{AS03} shifted to tailoring contact barriers of a proper shape. 

The second approach, that is central for the present paper, is based on the spin-valve transistor that is a three-terminal device operating on hot electrons.\cite{Mon98} This device supplements the perpendicular to plane geometry of GMR\cite{GMR} with semiconductor elements to achieve a large potential drop. Its recent modification including a tunnel barrier and called a magnetic tunnel transistor (MTT) source produces spectrally narrow spin-polarized electron beams. It already provided spin injection at the level of 10\% and magnetocurrents of about 3400 \%.\cite{SM01,Park1,Park2} 
In what follows I propose an analytical theory of such spintronic devices. This theory of three-terminal devices (like MTT), although based on a different physical concept, in its formalism generalizes the theory of two-terminal diffusive\cite{FZDS02} and ballistic\cite{KR03} devices and, therefore, establishes a connection to them. It is an important property of the MTT source that it employs both spin injection and spin filtering and, in this way, increases spin polarization of the electron beam. The theory demonstrates explicitly how these two mechanisms work together and specifies the parameters controlling spin polarized currents.

A basic scheme of MTT is shown in Fig.~1. A FM emitter is a source of a spin polarized electron current $j_e$. Because the transparency of a tunnel barrier decays fast, nearly exponentially, when electron energy $E$ decreases below the emitter Fermi energy $E_F^c$, the tunnel current is mostly confined to a narrow  region $E\approx E_F^e$, its width being restricted by the temperature $T$ and the $E$- dependence of the transparency. The device usually operates with $E$ just above the top of the Schottky barrier $E_S$, $E\agt E_S$,\cite{Park2} hence, the collector current $j_c$ originates from the spectrally narrow beam of ``active" electrons in the base, $E\approx E_F^e\approx E_S$. A textbook model of ballistic or quasielastically scattered electrons entering the collector inside its acceptance cone is hardly applicable to MTT. Indeed, (i) the scattering of fast electrons is usually inelastic, (ii) inelastic scattering knocks electrons out of the active-electron beam when its  spectral width is small, and (iii) electron scattering at the metallic interfaces is far from specular, hence, the notion of the acceptance angle is ambiguous. For these reasons, I employ a model in which inelastic scattering in the base plays the critical role, and $j_c$ comes only from the electrons that crossed the base without collisions. The rest of the electrons lose their energy and make the base current $j_b=j_e-j_c$.\cite{ballast} Because the scattering in the FM base is spin dependent, spin filtering in the base influences the spin polarization of $j_c$. Therefore, the spin polarization of the collector current depends both on the spin injection coefficient $\gamma_e$ of the FM emitter and on the spin filtering in the FM base. 

The model used below is based on the fact that when the active-electron energy is high enough over the Fermi energy of the base $E_F^b$, $V_{eb}\sim 1$ V, both electron-electron and electron-phonon scattering are inelastic and strongly enhanced in the magnitude. In particular, the data showing that the electron mean free path is usually much shorter for minority than for majority carriers\cite{Park2,RB00,comm} indicate an essential role of electron-electron scattering. Therefore, elastic scattering in the base will be disregarded. When $E_F^e\approx E_S$, the active-electron beam is spectrally narrow, and the very first inelastic collision experienced by an electron in the base removes it from the active group irreversibly if the temperature is low to moderate. One more simplification comes from the fact that the roughness of interfaces relaxes the in-plane momentum conservation essentially; indeed, wide angular distributions have been reported.\cite{SM01} Therefore, in what follows the interfaces are considered diffusive. This assumption simplifies the calculations significantly and reduces the number of arbitrary parameters. With these two basic assumptions, the system of coupled equations for MTT has been solved and explicit expressions for the basic parameters are presented.

The spin-transport problem should be solved in different parts of the device separately, and the solutions for spin-injection coefficients, $\gamma$, should be matched at the interfaces (in the spirit of the $\gamma$-technique of Ref.~\onlinecite{R02}). As usual, spin relaxation at the interfaces is neglected, i.e., they are assumed spin conserving.

{\it The FM emitter.} Inside the emitter, the current is carried mostly by the electrons in close vicinity of $E_F^e$. Therefore, the distribution function can be written as $f_\alpha(x,v_x)=f_0(E)+(df_0/dE)\psi_\alpha(x,v_x)$, where $\alpha=(\uparrow,\downarrow)$ is the spin index and $\psi_\alpha(x,v_x)$ depends only on the direction of the velocity $\bf v$. The Boltzmann problem for the emitter can be reduced to a Milne equation and solved when the spin-diffusion length $L_e$ is large as compared with the electron mean free path.\cite{KR03} For this purpose it is convenient to change from the functions $\psi_\alpha(x,v_x)$ to functions $e\zeta_\alpha(x,v_x)=-[\psi_\alpha(x,v_x)+e\varphi(x)]$, where $\varphi(x)$ is the electrostatic potential and $-e$, $(e>0)$, is the electron charge. Far from the tunnel junction, in the diffusion region, the functions $\zeta_\alpha(x,v_x)$ averaged over $\bf v$ acquire the meaning of electrochemical potentials $\zeta_\alpha(x)$. A solution of the Boltzmann problem depends on the boundary values $\zeta^<_\alpha$ of $\zeta_\alpha(x,v_x)$ for left moving electrons, $\zeta^<_\alpha$ being independent of $v_x<0$ because of the diffusive interface scattering. The spin injection coefficient $\gamma_e=(j_\uparrow-j_\downarrow)/j_e$ at the contact boundary, when expressed in terms of these constants, reads
\be
\gamma_e=(\zeta^<_\uparrow-\zeta^<_\downarrow)/2r_ej_e+\Delta r_e/r_e.
\label{eq1}
\ee
Here $j_e=j_\uparrow+j_\downarrow$ is the emitter current, resistances are
\be
r_e=(r_\downarrow^e+r_\uparrow^e)/4, ~~ \Delta r_e=(r_\downarrow^e-r_\uparrow^e)/4,~~r_\alpha^e=L_e/\sigma^e_\alpha,
\label{eq2}
\ee
$L_e$ is a spin-diffusion length, and $\sigma_\alpha^e$ are conductivities of spin-up and spin-down electrons.\cite{quarter}

{\it The tunnel barrier.} Spin polarized currents across a spin-conserving tunnel barrier can be written as
\be
j_\alpha=-e^2t_\alpha^{eb}\langle\zeta_\alpha v_\alpha\rangle_+ +t_\alpha^{be} j_\alpha^\leftarrow(L),
\label{eq3}
\ee
where brackets $\langle ... \rangle$ indicate averaging both over the direction of $\bf v$ and over energy, the subscript plus at the brackets shows that averaging over $\bf v$ should be performed only over the right hemisphere, and $t_\alpha^{eb}$ and $t_\alpha^{be}$ are the barrier transparencies for the right- and left-moving electrons, respectively, near the energy $E\approx E_F^e$. $j_\alpha^\leftarrow(L)$ stands for the current of those left-moving electrons at the left boundary of the base that were reflected elastically from the Schottky barrier and did not experience inelastic scattering in the base; hence, they have enough energy to return to the emitter. In Eq.~(\ref{eq3}), the currents of right- and left-moving electrons are written in a different form because the former electrons are described by a slightly perturbed Fermi distribution while the distribution of the latter ones is strongly nonequilibrium and the second term in Eq.~(\ref{eq3}) only includes active electrons, i.e. those that have not undergone inelastic scattering in the base.

Writing the currents $j_\alpha$ at both sides of the barrier through the contributions from the left- and right-moving electrons
\be
j_\alpha=-e^2[\langle\zeta_\alpha v_\alpha\rangle_+ +\zeta_\alpha^< \langle v_\alpha^e\rangle_-],~~j_\alpha=j_\alpha^\leftarrow(L)+j_\alpha^\rightarrow(L)
\label{eq4}
\ee
and eliminating the first terms of these expressions from Eqs.~(\ref{eq3}) and (\ref{eq4}), one finds
\be
(1-t_\alpha^{eb}-t_\alpha^{be})j_\alpha=
e^2t_\alpha^{eb}\zeta_\alpha^<\langle v_\alpha^e\rangle_- -t_\alpha^{be}j_\alpha^\rightarrow(L).
\label{eq5}
\ee
Using $\langle v_\alpha^e\rangle_-=-\rho_\alpha^ev_\alpha^e/4$, where $\rho_\alpha^e$ is the density of states at the emitter Fermi level, and applying the detailed balance principle $t_\alpha^{eb}\rho_\alpha^ev_\alpha^e=t_\alpha^{be}{\bar\rho}_\alpha^b{\bar v}_\alpha^b$, one finds the product $t_\alpha^{eb}\langle v_\alpha^e\rangle_-$ in terms of the parameters of the base. The bars in ${\bar\rho}_\alpha^b$ and ${\bar v}_\alpha^b$ indicate that these parameters of the electrons of the base should be taken at the Fermi energy of the emitter $E_F^e$ (rather than at the Fermi energy $E_F^b$ of the base). Therefore, ${\bar\rho}_\alpha^b$ and ${\bar v}_\alpha^b$ depend on the voltage $V_{eb}$, ${\bar\rho}_\alpha^b={\bar\rho}_\alpha^b(V_{eb})$ and ${\bar v}_\alpha^b={\bar v}_\alpha^b(V_{eb})$, and with $V_{eb}\sim 1$ V the difference may be rather strong. The product ${\bar \rho}_\alpha^b{\bar v}_\alpha^b$ can be related to the volume ${\bar V}_\alpha^b(k)$ contained inside the constant-energy surface $E_\alpha^b({\bf k})=E_F^e$ in the momentum space of the base as ${\bar\rho}_\alpha^b {\bar v}_\alpha^b=(1/8\pi^3\hbar)(d{\bar V}_\alpha^b/dk)$. Finally, Eq.~(\ref{eq5}) can be rewritten as
\be
j_\alpha=-[\zeta_\alpha^<+r_\alpha^bj_\alpha^\rightarrow(L)]/r_\alpha^t,
\label{eq6}
\ee
where
\be
r_\alpha^b=16\pi^2(h/e^2)dk/d{\bar V}_\alpha^b,~r_\alpha^t=r_\alpha^b(1-t_\alpha^{be}-t_\alpha^{eb})/t_\alpha^{be},
\label{eq7}
\ee
and $h=2\pi\hbar$. The currents $j_\alpha^\leftarrow(L)$ will be found below.

The resistances $r_\alpha^b=r_\alpha^b(V_{eb})$ depend only on the energy spectrum of the base and on the voltage $V_{eb}$. When $V_{eb}=0$, hence $E_F^b=E_F^e$, these resistances coincide with the Sharvin resistances of the base that are spin dependent because the base is ferromagnetic. Therefore, they can be considered as generalized Sharvin resistances.

The arguments similar to those of Ref.~\onlinecite{KR03} prove that the coefficients $r_\alpha^t$ are always positive, hence, they can be identified as spin dependent contact resistances.

Using Eq.~(\ref{eq6}) and the fact that the total current across the tunnel barrier equals $j_e$, one finds the spin injection coefficient across it
\bea
\gamma_t&=&\Delta r_t/r_t\nonumber\\
&-&\{(\zeta_\uparrow^<-\zeta_\downarrow^<)
+[r_\uparrow^bj_\uparrow^\rightarrow(L)-r_\downarrow^bj_\downarrow^\rightarrow(L)]\}/2r_tj_e,
\label{eq8}
\eea
where
\be
r_t=(r_\downarrow^t+r_\uparrow^t)/4,~~\Delta r_t=(r_\downarrow^t-r_\uparrow^t)/4.
\label{eq9}
\ee

{\it The base and the Schottky barrier.} Eq.~(\ref{eq6}) includes currents $j_\alpha^\rightarrow(L)$. To find them, the equations for the two right regions of Fig.~1 should be solved. The following assumptions will be made: (i) the scattering of active electrons in the base is inelastic, (ii) the scattering at the Schottky barrier is diffusive and elastic, and (iii) the scattering in the semiconductor to the right of the Schottky barrier is inelastic, hence, electrons lose their energy fast enough to not return to the base. E.g., if the barrier is steep (its shape is controlled by the potential $V_{cb}$) electrons acquire enough energy at a mean free path to emit optical phonons and lose energy fast enough. 

Under these conditions the current of right-moving electrons at the right edge of the base equals
\be
j_\alpha^\rightarrow(R)=E_2(d/l_\alpha)j_\alpha^\rightarrow(L),~
E_n(\xi)=\int_1^\infty {{dv}\over{v^n}} e^{-v\xi},
\label{eq10}
\ee
where $l_\alpha$ are the electron mean free paths in the base and $d$ is the base thickness. Functions $E_n(\xi)$ are typical of different transport problems. The function $E_2(\xi)\leq 1$, that decays faster than an exponent $\exp(-\xi)$, describes the decrease in the current of active electrons because of the inelastic scattering in the base. The enhanced decay comes from the cut-off of oblique electrons moving outside the cone that is getting narrower as $\theta_\alpha\sim\sqrt{l_\alpha/d}$ when $d/l_\alpha$ increases. With $t_\alpha^S$ being transparencies of the Schottky barrier for $\alpha$-electrons, the similar arguments result in the following expressions for the currents of left-moving active electrons at the left edge of the base
\be
j^\leftarrow_\alpha(L)=(t_\alpha^S-1)E_2^2(d/l_\alpha)j_\alpha^\rightarrow(L).
\label{eq11}
\ee
Eqs.~(\ref{eq4}) and (\ref{eq11}) permit one to find currents $j^\rightarrow_\alpha(L)$ in terms of the spin injection coefficient $\gamma_L$ at the left edge of the base and of the emitter current $j_e$
\be
j_\uparrow^\rightarrow(L)=(1+\gamma_L)j_e/2F_\uparrow,~
j_\downarrow^\rightarrow(L)=(1-\gamma_L)j_e/2F_\downarrow,
\label{eq12}
\ee
where the factors
\be
F_\alpha=[1-(1-t_\alpha^S)E_2^2(d/l_\alpha)]\leq 1
\label{eq13}
\ee
describe spin filtering in the FM base and by the Schottky barrier. Substituting Eqs.~(\ref{eq12}) into Eq.~(\ref{eq8}), one finds that spin-dependent resistances of the base $r_\alpha^b$ combine with filtering factors $F_\alpha$. Their ratios
\be
r_\alpha^f=r_\alpha^b/F_\alpha
\label{eq14}
\ee
acquire the meaning of new effective resistances of the base.

{\it Spin injection coefficients.} Using Eqs.~(\ref{eq1}), (\ref{eq8}), and (\ref{eq12}), one can eliminate the factor $(\zeta^<_\uparrow-\zeta^<_\downarrow)$ and the currents $j_\alpha^\rightarrow$. Then, applying the spin-current conservation condition, $\gamma_e=\gamma_t=\gamma_L$, one finds the spin injection coefficient of the emitter 
\be
\gamma_e=(\Delta r_e+\Delta r_t+\Delta r_f)/(r_e+r_t+r_f),
\label{eq15}
\ee
where
\be
r_f=(r_\downarrow^f+r_\uparrow^f)/4,~~\Delta r_f=(r_\downarrow^f-r_\uparrow^f)/4.
\label{eq16}
\ee
Eq.~(\ref{eq15}) resembles the equations of the diffusive\cite{R02} and ballistic\cite{KR03} theory for spin injection coefficients. In all cases, the denominator of $\gamma$ includes a sum of the effective resistances of all elements while the numerator includes a sum of their spin selectivities, i.e., the differences of the effective resistances for spin-down and spin-up electrons. Of course, specific expressions for these resistances depend on the system and on the model.

Because MTT is considered as a source for spin injection  into a semiconductor collector across a Schottky barrier, the spin injection coefficient $\gamma_c=(j_\uparrow^c-j_\downarrow^c)/j_c$ is of primary interest. In the framework of our model $j_\alpha^c=t_\alpha^Sj_\alpha^\rightarrow(R)$, hence, one easily finds
\be
\gamma_c=({\cal F}_\downarrow r_\downarrow-{\cal F}_\uparrow r_\uparrow)/
({\cal F}_\downarrow r_\downarrow+{\cal F}_\uparrow r_\uparrow)
\label{eq17}
\ee
where
\be
r_\alpha=r_\alpha^e+r_\alpha^t+r_\alpha^f,~~
{\cal F}_\alpha=F_\alpha/t_\alpha^SE_2(d/l_\alpha).
\label{eq18}
\ee
Comparing Eqs.~(\ref{eq15}) and (\ref{eq17}) shows that both expressions include the same total effective resistances $r_\alpha$ of the spin channels. However, Eq.~(\ref{eq15}) includes their differences and sums directly while in Eq.~(\ref{eq17}) these resistances are weighted with the factors ${\cal F}_\alpha$.

In the same notations, the transfer ratio of the transistor equals
\be
j_c/j_e=(r_\downarrow/{\cal F}_\uparrow+r_\uparrow/{\cal F}_\downarrow)/(r_\downarrow+r_\uparrow).
\label{eq19}
\ee
Eqs.~(\ref{eq17}) and (\ref{eq19}) are the final results of the theory.

{\it Discussion.} The above simple theory of a MTT spin-polarized electron source is based on the assumption that the collector current is carried by collisionless active electrons injected into the base from the emitter in a narrow spectral region near its Fermi energy $E_F^e$. Elastic scattering in the base was disregarded while inelastic scattering, removing electrons irreversibly from the active region, confines the active electrons inside a central cone even when interfaces are diffusive.

Eq.~(\ref{eq17}) for the spin injection coefficient of the MTT source reflects the interplay of the spin-emission and spin-filtering cascades (the emitter + tunnel contact block and the base + Schottky contact block, respectively). The former cascade is described by the resistances $r_\alpha^{et}=r_\alpha^e+r_\alpha^t$. Its efficiency as a spin polarizer and the sign of the spin polarization depend of the difference $r_\downarrow^{et}-r_\uparrow^{et}$. The latter cascade is described by generalized Sharvin resistances $r_\alpha^b$ and filtering coefficients $F_\alpha$ and ${\cal F}_\alpha$. It is seen from Eqs.~(\ref{eq14}), (\ref{eq17}) and (\ref{eq18}) that these coefficients enter into $\gamma_c$ as weighting factors renormalizing the resistances $r_\alpha$ and $r_\alpha^b$. A direct calculation shows that $d\gamma_c/dt_\uparrow^S>0$ and $d\gamma_c/dl_\uparrow>0$. Therefore, an increase in $t_\uparrow^S$ and $l_\uparrow$ enhances or reduces spin polarization of the source depending on the sign of the spin polarization provided by the first cascade.

Resistances $r_\alpha$ in Eq.~(\ref{eq17}) are sums of the resistances of different elements of the device. Therefore, $\gamma_c$ is controlled by the elements having the largest resistances. In particular, ballistic transport in the base cannot solve the conductivity mismatch problem.\cite{KR03} When the scattering in the base is weak, there are its Sharvin resistances $r_\alpha^f$ (modified by the filtering) that compete with the resistances of the different elements of the device.

MTT operates with the energy of hot electrons $E\approx E_S$. When $E>E_S$, tunnel transparencies $t_\alpha^S$ decrease with $(E-E_S)$ according to a power law because of the potential singularity at the interface. Energy dependence of $\gamma_c$ and $j_c/j_e$ in this region is sensitive to the barrier shape. When $E$ is essentially larger than $E_S$, the theory is not applicable because of increasing number of electrons that underwent inelastic scattering but still retained enough energy to overcome the Schottky barrier. 

The effect of the elastic scattering in the base and the specular component of the interface scattering that have been disregarded are both material and technology dependent. The contributions of these mechanisms cannot be found from general arguments and should be evaluated for specific devices from the experimental data.

In conclusion, a theory of a magnetic tunnel transistor source of spin polarized electrons based on the concept of the dominant role of inelastic electron scattering has been developed. The theory takes into account the coupling between the spin-injection and spin-filtering blocks of the device. Explicit expressions for the spin injection coefficient $\gamma_c$ and the transistor transfer ratio $j_c/j_e$ are presented.

The financial support from DARPA/SPINS by the ONR Grant N000140010819 is gratefully acknowledged.

\begin{figure}[th]
%\vskip 0.2truecm
%\begin{center}
%\epsfig{file=figNewW.eps, angle=0, width=0.4\textwidth}
%\end{center}
%\vskip-0.5truecm
\caption{Schematic energy band diagram of a magnetic tunnel transistor source. From left to right: a metallic FM emitter, an insulating tunnel barrier, a metallic FM base, and a semiconductor collector separated from the base by a Schottky barrier. The difference between the Fermi energies of the emitter and the base, $E^e_F$ and $E^b_F$, is controlled by the voltage $V_{eb}$. $E^e_F$ nearly coincides with the top of the Schottky barrier when spin injection across it reaches its maximum. Energy distribution of the electrons emitted across the tunnel barrier is shown schematically.}
%\label{angular}
\end{figure}
\vskip-0.0truecm 

\newpage
\epsfig{file=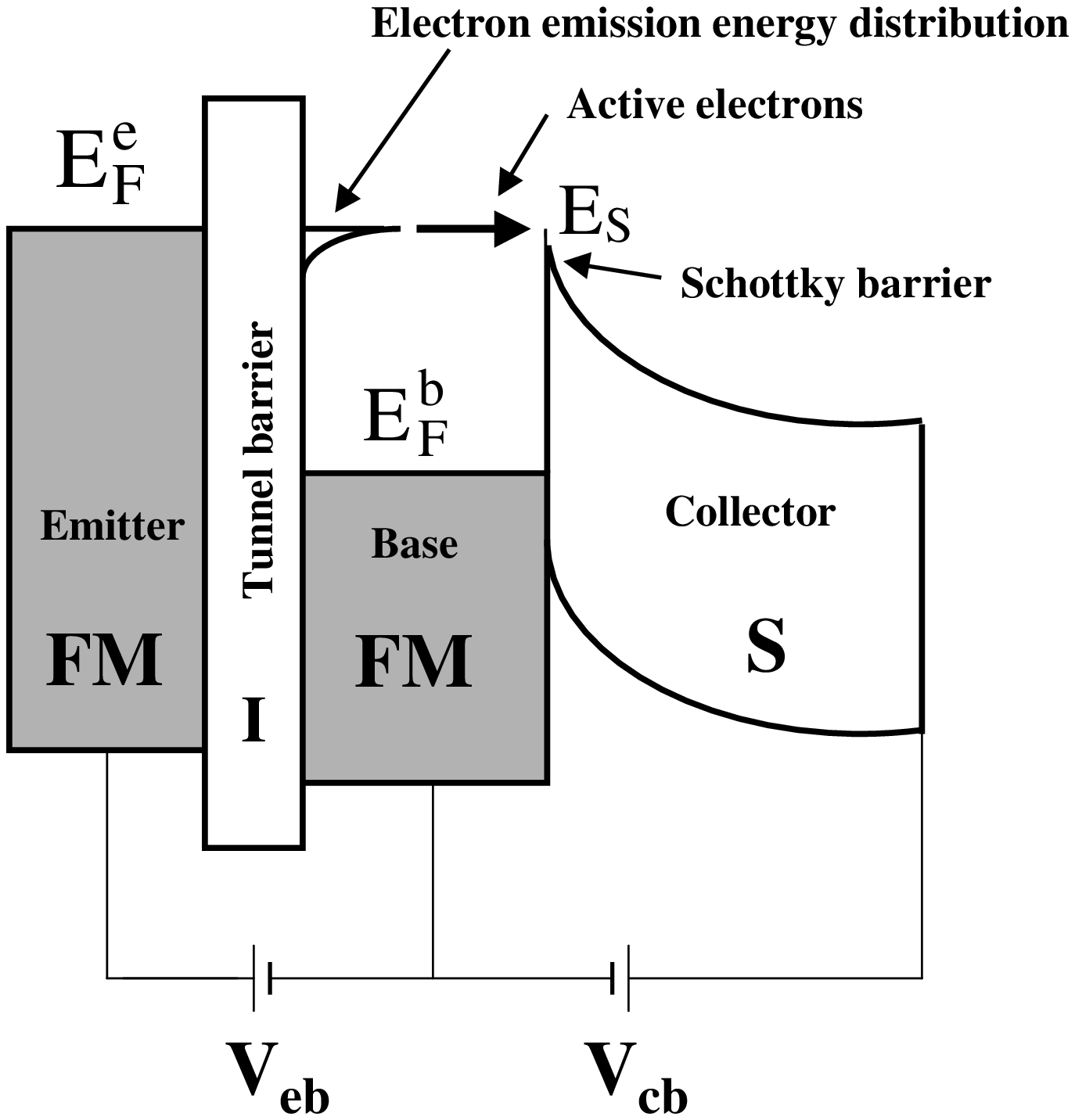,angle=-90}


\begin{thebibliography}{99}
\bibitem[*]{Rashba*} Email: erashba@mailaps.org
\bibitem{reviews} S. A. Wolf, D. D. Awschalom, R. A. Buhrman, J. M. Daughton, S. von Molnar, M. L. Roukes, A. Y. Chtchelkanova, and D. M. Treger, Science {\bf 294}, 1488 (2001).
\bibitem{JS87} M. Johnson and R. H. Silsbee, Phys. Rev. {\bf 35}, 4959 (1987), Appendix. 
\bibitem{Sch00}G. Schmidt, D. Ferrand, L. W. Molenkamp, A. T. Filip, B. J. van Wees, Phys. Rev. B {\bf 62}, R4790 (2000).
\bibitem{R00} E. I. Rashba, Phys. Rev. B {\bf 62}, R16267 (2000).
\bibitem{Plo01} H. J. Zhu, M. Ramsteiner, H. Kostial, M. Wassermeier, 
H.-P. Sch\"onherr, and K. H. Ploog, Phys. Rev. Lett.  {\bf 87}, 016601 (2001).
\bibitem{HJ01} P. R. Hammar and M. Johnson, Appl. Phys. Lett. {\bf 79}, 2591 (2001).
\bibitem{Jon01} A. T. Hanbicki, B. T. Jonker, G. Itskos, G. Kioseoglou, and A. Petrou, Appl. Phys. Lett. {\bf 80}, 1240 (2002).
\bibitem{Saf01} V. F. Motsnyi, V. I. Safarov, J. De Boeck, J. Das, W. Van Roy, E. Goovaerts, G. Borghs, Appl. Phys. Lett. {\bf 81}, 265 (2002).
\bibitem{Weiss02} S. Kreuzer, J. Moser, W. Wegscheider, D. Weiss, M. Bichler, and D. Schuh, Appl. Phys. Lett. {\bf 80}, 4582 (2002).
\bibitem{Jap02} C.-M. Hu, J. Nitta, A. Jensen, J. B. Hansen, H. Takayanagi, T. Matsuyama, D. Heitmann, and U. Merkt, J. Appl. Phys. {\bf 91}, 7251 (2002).
\bibitem{Samarth02} S. H. Chun, S. J. Potashnik, K. C. Ku, P. Schiffer, and N. Samarth, Phys. Rev. B {\bf 66}, 100408 (2002).
\bibitem{Dutch02} F. J. Jedema, H. B. Heersche, A. T. Filip, J. J. A. Baselmans, and van Wees, {\it Nature} {\bf 416} (6882), 713 (2002).
\bibitem{Han03} A. T. Hanbicki, O. M. J. van 't Erve, R. Mango, G. Kioseoglou, C. H. Li, G. Itskos, R. Mallori, M. Yaser, and A. Petrou, Appl. Phys. Lett. {\bf 82}, 4092 (2003).
\bibitem{AS03} J. D. Albrecht and D. L. Smith, Phys. Rev. B {\bf 68}, 035340 (2003).
\bibitem{Mon98} D. J. Monsma, R. Vlutters, and J. C. Lodder, Science {\bf 281}, 407 (1998).
\bibitem{SM01} R. Sato and K. Mizushima, Appl. Phys. Lett. {\bf 79}, 1157 (2001).
\bibitem{GMR} M. N. Baibich, J. M. Broto, A. Fert, F. Nguyen Van Dau, F. Petroff, P. Etienne, G. Creuzet, A. Friederich, and J. Chazelas, Phys. Rev. Lett. {\bf 61}, 2472 (1988).
\bibitem{Park1} X. Jiang, R. Wang, S. van Dijken, R. Shelby, R. Macfarlane, G. S. Solomon, J. Harris, and S. S. P. Parkin, Phys. Rev. Lett. {\bf 90}, 256603 (2003).
\bibitem{Park2} S. van Dijken, X. Jiang, and S. S. P. Parkin, Appl. Phys. Lett. {\bf 83}, 951 (2003).
\bibitem{vS87} P. C. van Son, H. Kempen, and P. Wyder, Phys. Rev. Lett. {\bf 58}, 227 (1987).
\bibitem{HZ97} S. Hershfield and H. L. Zhao, Phys. Rev. B {\bf 56}, 3296 (1997).
\bibitem{R02} E. I. Rashba, Euro. Phys. J. B {\bf 24}, 513 (2002).
\bibitem{FZDS02} J. Fabian, I. \v{Z}uti\'{c}, and S. Das Sarma, Phys. Rev. {\bf 66}, 165301 (2002).
\bibitem{KR03} V. Ya. Kravchenko and E. I. Rashba, Phys. Rev. B {\bf 67}, 121310(R) (2003).
\bibitem{quarter} Factor $1/4$ ensures that the resistance $r_e$ equals $1/2$ of the resistance of each of the spin channels when $r^e_\uparrow=r^e_\downarrow$.
\bibitem{ballast} Here and below $j_e$ includes the contribution of the active electrons only. Parasitic currents coming from the resonant tunneling at lower energies are disregarded.
\bibitem{RB00} W. H. Rippard and R. A. Buhrman, Phys. Rev. Lett. {\bf 84}, 971 (2000).
\bibitem{comm} The notion of the minority and majority electrons is not unambiguous and may depend on the definition (the contribution to the magnetization or to specific transport processes). This uncertainty is immaterial for the phenomenological theory of this paper.


\end{thebibliography}
\end{document}